\pdfoutput=1
\documentclass[11pt]{article}

\usepackage[T1]{fontenc}
\usepackage{lmodern}
\usepackage[margin=1in]{geometry}
\usepackage{cite}
\usepackage{amsmath,amssymb,amsfonts}
\usepackage{graphicx}
\usepackage{textcomp}
\usepackage{booktabs}
\usepackage{multirow}
\usepackage{tabularx}
\usepackage{xcolor}
\usepackage{listings}
\usepackage{xurl}
\usepackage{authblk}
\usepackage[colorlinks=true,linkcolor=blue!50!black,citecolor=blue!50!black,urlcolor=blue!50!black]{hyperref}
\usepackage[capitalise,noabbrev]{cleveref}
\crefname{figure}{Fig.}{Figs.}
\Crefname{figure}{Fig.}{Figs.}
\crefname{section}{Section}{Sections}
\Crefname{section}{Section}{Sections}
\crefname{subsection}{Section}{Sections}
\Crefname{subsection}{Section}{Sections}
\DeclareRobustCommand{\mdl}[1]{\path{#1}}  
\newcommand{\cond}[1]{\textsf{#1}}

\newsavebox{\answerboxbin}
\newenvironment{answerbox}
  {\par\smallskip\noindent\setlength{\fboxrule}{0.4pt}\setlength{\fboxsep}{4pt}%
   \begin{lrbox}{\answerboxbin}%
   \begin{minipage}{\dimexpr\linewidth-2\fboxsep-2\fboxrule\relax}\setlength{\parindent}{0pt}}
  {\end{minipage}\end{lrbox}%
   \fcolorbox{gray!40}{gray!6}{\usebox{\answerboxbin}}\par\smallskip}

\title{Trusting the Tool: Asymmetric Deference of LLM Vulnerability Auditors to Static-Analysis Evidence}

\author[1,2]{ZhiHong Liu}
\author[1]{Shuyao Gao\thanks{Corresponding author. E-mail: \texttt{gaoshuyao.mk@gmail.com}}}
\affil[1]{Seoul School of Integrated Sciences and Technologies (aSSIST University), Seoul, Republic of Korea}
\affil[2]{Liaoning Mechanical \& Electrical College of Technology, Dandong, Liaoning, China}
\date{}

\begin{document}

\maketitle

\begingroup
\renewcommand\thefootnote{}
\footnotetext{This work has been submitted to the IEEE for possible publication. Copyright may be transferred
without notice, after which this version may no longer be accessible.}
\addtocounter{footnote}{-1}
\endgroup

\begin{abstract}
Pipelines that combine large language models (LLMs) with static analysis let the model consume the analyser's output
as evidence, but whether the model distinguishes true from false evidence, and whether its scrutiny depends on what
the evidence claims, has not been measured. We built an intraprocedural analyser for C/C++ that produces proof
obligations and audited vulnerable and patched functions from PrimeVul under six evidence conditions that differ only
in the notes, including fabricated assurances that a missing check exists and fabricated alarms that an added check is
missing. The study covers 916 pairs in four disjoint sets, eleven model configurations from four vendors and
16{,}572 audits; no model is trained or fine-tuned. True evidence barely improved accuracy ($+0.02$). False evidence
acted asymmetrically: relative to true notes about the same operations, false assurances lowered the detection of real
vulnerabilities by up to 19 points, whereas false alarms never significantly lowered accuracy and were disputed far
more often. An internally registered replication on seven new models
confirmed that false assurances move verdicts towards their claim more than false alarms do (mean asymmetry $+0.16$,
95\% CI $+0.10$ to $+0.24$). Instructions in the audit prompt did not reliably reduce deference; the best one failed
to replicate. Verifying each claim in a separate call and showing the refuted claims raised accuracy under false
assurances by 12 points in a pre-specified test ($p=0.002$) without loss under true evidence; for a second model the
gain was smaller and not significant. Pipelines should therefore verify a tool's guarantees before the model relies on
them, and studies should measure deference in a direction-aware way.
\end{abstract}


\section{Introduction}
\label{sec:intro}
Large language models (LLMs) are increasingly used to audit code for vulnerabilities, and a natural way
to make them more reliable is to give them the output of program analysis. Systems that combine LLMs with static
analysis let the analyser enumerate candidate defects and ask the model to judge them~\cite{li2024llift,li2025iris},
or let the model triage the analyser's warnings~\cite{li2023assisting,du2026reducing}. In all of these designs the
model consumes evidence produced by a tool, and the value of the combination depends on how the model treats that
evidence: whether it uses true evidence, and whether it notices when the evidence is wrong. Static analysers are
wrong often enough that developers cite false positives as a main reason not to use them~\cite{johnson2013why},
and the same analysers can also be wrong in the other direction, reporting that a guard exists or that a path is
safe when it is not.

How people treat the output of automated aids has been studied for decades under the name of automation bias:
people over-rely on automation, accept its errors of commission, and miss what it fails to flag
\cite{parasuraman1997humans,goddard2012automation}. Recent work in human--AI interaction shows the same pattern with
AI assistants, where explanations can increase rather than reduce over-reliance~\cite{bucinca2021trust} and fluent
assistance can create confidence without competence~\cite{eleftheriou2026confidence}. LLMs that consume tool output
are in the position of the human user. Work on knowledge conflicts shows that LLMs can be highly receptive to
external evidence that contradicts what they know~\cite{xie2024chameleon,xu2024conflicts,wu2024clasheval}, and
work on sycophancy shows that they tend to agree with the views put to them~\cite{sharma2024sycophancy}. Concurrent
work has begun to show that context outside the code, such as prior verdicts, pull-request metadata and comments,
shifts the verdicts of LLM security reviewers~\cite{shahriar2026words,alexopoulos2026contextual,wu2026alibi}. These
studies vary cues that are not tied to the truth of a specific claim, or study only claims that the code is safe.
The closest, ALIBI~\cite{wu2026alibi}, already embeds fabricated tool reports in code comments to evade LLM
detectors, but it studies only claims of safety and does not compare a false claim with a true claim about the same
operation. Our study adds three things: true and false evidence matched on the same operations, false claims in both
directions, and mitigations evaluated on held-out data.
Whether LLM auditors distinguish true from false analysis evidence about the same operations, and whether their
scrutiny depends on the direction of the claim, has not been measured.

This paper reports a controlled empirical study of that question. We built an intraprocedural analyser for C and
C++ that produces \emph{proof obligations}: for each sensitive operation, the guards that hold on every path to it
and what the function does not guarantee. Running it on pairs of vulnerable and patched functions from
PrimeVul~\cite{ding2024primevul}, we audit the same function under six evidence conditions that vary only the
notes: none, all true notes, true notes about the operations the fix touches, true notes about other operations, a
fabricated \emph{assurance} that the missing check exists, and a fabricated \emph{alarm} that the check added by the
fix is missing. Because both false conditions are compared with true notes about the same operations, differences
in the verdicts can be attributed to the truth of the evidence. All models are used through their APIs for
inference only; no model training or fine-tuning is performed. We ask four research questions: whether true
evidence helps (RQ1), how relevance and correctness affect the verdicts (RQ2), whether the models check the
evidence (RQ3), and whether the way evidence is presented can reduce deference to false evidence (RQ4).

The main findings are as follows.
\begin{itemize}
\item \emph{True evidence barely helps.} Showing all true notes changes accuracy by $+0.02$ (95\% CI $-0.03$ to
  $+0.07$); restricting them to the relevant operations gives a small, non-significant gain (RQ1).
\item \emph{False evidence acts asymmetrically.} A fabricated assurance lowers the detection of real vulnerabilities
  (by 0.11, $p=0.012$, for \mdl{deepseek-chat}, and by 0.13--0.19 for four of seven further models), whereas a
  fabricated alarm about a fixed function never significantly lowers accuracy and often raises it, because the model goes looking
  for the guard and finds it (RQ2). Models dispute false alarms in the text of their answers far more often than
  false assurances (RQ3).
\item \emph{The asymmetry generalises.} Accuracy contrasts make deference look model-dependent, because models that
  flag nearly every function cannot be pushed further. A direction-aware measure reduces this confound, although its size still depends on how much room a model's
  leaning leaves in each direction. After deriving it on four configurations, we registered a replication internally on seven new models from four vendors, which
  confirmed it: false assurances move verdicts towards their claim more than false alarms do (mean asymmetry
  $+0.16$, 95\% CI $+0.10$ to $+0.24$, $p<0.001$).
\item \emph{Separate verification mitigates deference; instructions do not reliably.} The best instruction-level
  strategy on the study set (reframing assurances as claims to verify, $+0.18$) failed to replicate on held-out data.
  Checking each note against the code in a separate call and showing the refuted notes (verify-show) recovered 12
  points of accuracy under false assurances in a pre-specified test on 300 unseen pairs ($p=0.002$) at no cost under
  true evidence; for a second model the gain was half as large and not significant.
\end{itemize}

For practitioners who build LLM--analysis pipelines, these findings point to a concrete risk: the input that
most needs checking is the analyser's guarantees, not its alarms, because guarantees are the evidence the models
accept without looking. We return to this point in \cref{sec:implications} and in the conclusion.

The study is designed to support these claims with appropriate rigour. It comprises 916 PrimeVul pairs (1{,}832
functions) in four disjoint sets, eleven model configurations from four vendors, and 16{,}572 audits. All
comparisons are paired on the same functions. Exploration and confirmation are separated: presentation strategies
were developed on a development set, and the hypotheses of the confirmatory phases were fixed before the corresponding
data were collected~\cite{nosek2018preregistration,ernst2023registered}; \cref{sec:protocol} states how each was
recorded. The behavioural coding is validated by two
independent LLM annotators and a human check. We report every strategy we tried and every confirmatory test,
including those that failed.

The main contributions of this study are:
\begin{enumerate}
\item \textbf{A controlled methodology for measuring how LLMs use program-analysis evidence.} It consists of an
  obligation analyser for C/C++, six evidence conditions built from vulnerability-fixing pairs, matched contrasts
  that isolate the truth of the evidence, and a direction-aware measure of deference relative to true evidence about the same
  operations.
\item \textbf{Empirical evidence of asymmetric deference.} Across eleven configurations and four vendors, LLM
  auditors trust false assurances and check false alarms. The asymmetry was confirmed in an internally registered
  replication; its size varies between models and samples, and the study is limited to C/C++ functions audited in isolation.
\item \textbf{An evaluation of evidence presentation.} We evaluate eight ways of presenting evidence along five
  dimensions (content, framing, provenance, checking and timing), including a failed replication, and a pre-specified
  test showing that moving verification into a separate step reduces deference without losing the value of true
  evidence for one of two models.
\item \textbf{A replication package.} It contains the analyser, the evidence of every condition, all model responses,
  the annotations, the registrations and the analysis scripts that produce every number and figure in this
  paper.\footnote{\url{https://github.com/lzh0082-cloud/CODE-DETECTION}}
\end{enumerate}

The remainder of the paper is organised as follows. \Cref{sec:related} discusses related work.
\Cref{sec:preliminary} reports a preliminary study that motivated the design. \Cref{sec:method} describes the
methodology and \cref{sec:setup} the experimental setup, \cref{sec:results} the results, and \cref{sec:discussion}
their implications and limitations. \Cref{sec:conclusion} concludes.

\section{Related Work}
\label{sec:related}
Our study sits at the intersection of six strands of prior work. We summarise each briefly and indicate where the
present paper departs from them.

\subsection{LLMs for Vulnerability Detection}

LLMs are now applied across software engineering tasks~\cite{fan2023llmse,hou2024llm4se}, and vulnerability
detection is one of the most studied. Before LLMs, deep learning detectors learned from token sequences, code
gadgets or program graphs~\cite{li2018vuldeepecker,zhou2019devign,fu2022linevul}, trained on datasets mined from
vulnerability-fixing commits~\cite{fan2020bigvul,chen2023diversevul}. Such detectors have repeatedly been found to
perform far worse under realistic evaluation than on their original benchmarks~\cite{chakraborty2022deep}. Risse and
B\"ohme showed that state-of-the-art models overfit to features unrelated to the vulnerability and cannot
distinguish vulnerable functions from their patched versions~\cite{risse2024limits}. PrimeVul~\cite{ding2024primevul}
addressed label noise and data duplication and introduced paired evaluation; on it, code language models, including
large general-purpose LLMs, rarely judge both versions of a pair correctly. Studies focusing on LLMs report the same
picture: answers are non-deterministic, reasoning is often incorrect or unfaithful, and small semantics-preserving
changes alter verdicts~\cite{ullah2024llms}. Prompting strategies and in-context information improve results only
modestly~\cite{zhou2024nier,khare2025understanding}. Steenhoek et al.\ argue that vulnerability detection requires
multi-step reasoning about code semantics, such as relating bounds and null checks to the operations they protect,
and that this reasoning is where LLMs fail~\cite{steenhoek2024err}. Our preliminary study (\cref{sec:preliminary})
agrees: the models we study make almost no errors about local program structure, so their errors are judgements,
not misreadings. We use PrimeVul's paired data and its paired metrics, but our dependent variable is not detection
performance; it is how detection changes when the model is given evidence.

\subsection{Combining LLMs and Static Analysis}

Static analysers trade completeness for precision in ways that limit their adoption. They miss many real
bugs~\cite{habib2018many}, and their false positives are a main reason why developers ignore or abandon
them~\cite{johnson2013why,christakis2016developers,sadowski2018lessons}. A body of work therefore post-processes
static-analysis alarms to identify the actionable ones~\cite{heckman2011systematic}, although learned alarm
classifiers have been found to benefit from data leakage in their evaluation~\cite{kang2022detecting}. LLMs offer a
new way to perform this triage, and more generally to use tools as part of their
reasoning~\cite{schick2023toolformer,yao2023react}. A growing line of work combines LLMs with static analysis.
LLift uses an LLM to examine the candidate bugs of a static analyser for the Linux kernel and to discard those whose
preconditions are ruled out~\cite{li2024llift}; an earlier pilot used ChatGPT to prune the analyser's false
positives~\cite{li2023assisting}. IRIS lets an LLM infer taint specifications that a static analyser then uses for
whole-repository detection~\cite{li2025iris}. In industry, hybrid pipelines in which an LLM adjudicates
static-analysis alarms remove most false positives while keeping high recall~\cite{du2026reducing}, and
general-purpose coding agents can filter most false positives of security analysers, at the risk of also
suppressing real vulnerabilities~\cite{xiong2026sifting}. In all of these designs, analysis results enter the
model's context as evidence. Their evaluations measure the end result of the pipeline; they do not isolate how the
model treats correct and incorrect evidence. Our study complements them: we hold the code and the model fixed and
vary only the truth, relevance and presentation of the evidence.

\subsection{LLMs and External Evidence}

Natural language processing research has studied how LLMs resolve conflicts between their parametric knowledge and
evidence in their context. LLMs can be highly receptive to coherent external evidence even when it contradicts what
they know, while also showing confirmation bias towards evidence that agrees with
them~\cite{xie2024chameleon}; the literature on such knowledge conflicts is surveyed by Xu et
al.~\cite{xu2024conflicts}. ClashEval found that models adopt incorrect retrieved content over their own correct
answer in more than 60\% of cases, and less often the more implausible the error is~\cite{wu2024clasheval}. A
related tendency, sycophancy, makes models agree with the views of the
user~\cite{perez2023discovering,sharma2024sycophancy}, and models often abandon a correct answer when it is
challenged, even by invalid arguments~\cite{wang2023defend}. These studies vary the content of factual evidence in
question answering. We study evidence about program behaviour, which the model can check against the code it is
given, and we find that \emph{the direction} of the evidence (whether it says that the code is safe or that it is
unsafe) determines whether it is checked.

\subsection{Contextual Bias in LLM Security Review}

Concurrent work shows that LLM security verdicts depend on context outside the code. Shahriar et al.\ add author
attributions, task framings and prior analysis outcomes (a function ``marked SAFE'' or ``VULNERABLE'' in an
automated analysis) to PrimeVul functions and find that all eight models they study are susceptible; they conclude
that such cues mainly change how willing a model is to flag code rather than how well it
discriminates~\cite{shahriar2026words}. Alexopoulos et al.\ show that crafted pull-request metadata biases
LLM-based code review and, with iterative refinement, hides 32 of 33 real vulnerabilities from production
pipelines, while blunt templates can backfire because they raise suspicion~\cite{alexopoulos2026contextual}.
ALIBI uses code comments with fabricated safety claims, including fabricated tool reports, to evade LLM detectors,
and finds that isolating the code from unverifiable comments defends better than instructions to distrust
them~\cite{wu2026alibi}. In contrast, Thornton finds that adversarial comments have no significant effect on
frontier models in a benchmark of 100 samples~\cite{thornton2026comments}. Beyond verdicts, correct
vulnerability verdicts are often accompanied by fabricated or unverifiable reasoning~\cite{hu2026vera}, and in code
smell detection, models adopt false premises in the prompt unless asked to reason from evidence
first~\cite{fahad2026sycophancy}.

\subsection{Hallucination, Reasoning and Verification}

LLMs produce fluent statements that are unsupported by their input or by facts, a behaviour broadly termed
hallucination~\cite{ji2023survey,huang2025hallucination}. Generating a chain of reasoning improves multi-step
reasoning~\cite{wei2022cot}, but the stated reasoning need not reflect what determines the
answer~\cite{turpin2023unfaithful}. Whether models can correct their own errors is contested: iterative
self-feedback improves some tasks~\cite{madaan2023selfrefine}, but without external feedback models generally
fail to correct their reasoning and may make it worse~\cite{huang2024selfcorrect}, and self-correction works best
when reliable external feedback is available~\cite{kamoi2024selfcorrection}. Chain-of-Verification reduces
hallucination by having the model plan verification questions about its draft, answer them independently and revise
the draft~\cite{dhuliawala2024cove}. Our verify-show strategy follows the same principle of independent
verification, applied to the tool's claims rather than to the model's own draft; the code under audit serves as the
external reference against which the claims are checked.

\subsection{Automation Bias and Cognitive Forcing}

Human factors research distinguishes misuse of automation, that is over-reliance, from disuse, which is often caused
by false alarms~\cite{parasuraman1997humans}; appropriate reliance requires trust to be calibrated to the actual
reliability of the aid~\cite{lee2004trust}. Automation bias produces two kinds of error: errors of commission, when
people follow incorrect advice, and errors of omission, when they miss problems the aid did not
flag~\cite{skitka1999automation}. A systematic review of automation bias in clinical decision support found that
incorrect advice leads to errors of commission, and identified the presentation of advice and the user's
verification effort as mediators~\cite{goddard2012automation}. In AI-assisted decision making, explanations can
increase over-reliance~\cite{bansal2021whole}; they reduce it when verifying the AI's output is easy relative to
solving the task alone~\cite{vasconcelos2023explanations}. Cognitive forcing functions, such as asking people to
decide before they see the AI's recommendation, also reduce over-reliance~\cite{bucinca2021trust}. Recent work on
LLM-assisted knowledge work finds that fluent assistance produces high perceived understanding with the lowest
actual learning~\cite{eleftheriou2026confidence}. We transfer these concepts to an LLM that is the user of a tool:
we ask whether the model over-relies on the tool's assurances, and we test interventions analogous to those studied
with people, namely framing, provenance, independent verification and a commit-first variant of cognitive forcing.

\subsection{Pre-Registration in Software Engineering}

Empirical software engineering has long-standing guidelines for controlled
experiments~\cite{wohlin2012experimentation} and for the statistical analysis of stochastic
results~\cite{arcuri2011practical}. LLM-based studies add threats of their own: closed models change behind their
APIs, training data may overlap with evaluation data, and outputs are non-deterministic even at temperature
zero~\cite{sallou2024breaking,ouyang2025nondeterminism}, while results can depend on seemingly irrelevant details of
prompt formatting~\cite{sclar2024quantifying}. Pre-registration commits to hypotheses and analyses before the data
are seen, which separates confirmatory from exploratory findings and prevents selective
reporting~\cite{nosek2018preregistration}. It has been adopted by software engineering venues in the form of
registered reports~\cite{ernst2023registered}. Studies of LLMs are especially exposed to the garden of forking
paths, because prompts can be varied without limit; we therefore separate development and test data and fix the
hypotheses of the confirmatory phases in advance.

\subsection{Summary of the Gap}

Work on LLM vulnerability detection measures how well models flag code, not how they use evidence; work that
combines LLMs with static analysis evaluates whole pipelines, not the model's treatment of correct and incorrect
evidence; and work on knowledge conflicts studies factual evidence in question answering that the model cannot check
against its input. Concurrent studies of contextual bias in security review come closest, and our study differs from
them in three ways. First, our evidence consists of specific claims about the operations in the function, which the
model can check against the code, and each false claim is compared with a true claim about the same operation; this
isolates the truth of the evidence from its presence. Second, we compare false claims in both directions on matched
functions and show that the models scrutinise them differently: they check alarms and trust assurances. A pure
change in willingness to flag, as reported for generic cues~\cite{shahriar2026words}, would move verdicts by similar
amounts in both directions; our direction-aware measure shows that it does not. Third, we evaluate mitigations in a
pre-specified test on held-out data. The adversarial studies above examine only claims that code is safe; our results
explain why that is the direction an attacker would choose. The asymmetry also bears directly on alarm triage,
where the model is asked to dismiss alarms, and on designs that pass the analyser's guarantees to the model as
context.

\section{Preliminary Study: Where Do the Errors Come From?}
\label{sec:preliminary}

The design of the main study grew out of two preliminary experiments with \mdl{deepseek-chat}. The first asked
whether LLMs misread program structure, which static analysis could correct; the second whether static-analysis
facts improve vulnerability audits.

\subsection{Claim-Level Errors in Execution Reasoning}
We asked the model to predict the output of Python programs from CRUXEval~\cite{gu2024cruxeval} (50 samples) and
from the HumanEval part of REval~\cite{chen2025reval} (85 samples), reasoning step by step. We extracted the factual
claims of each reasoning chain into a small predicate language (for example, that a line executes, that a branch is
taken, how often a loop iterates or what value a variable holds) and labelled each claim against the program's
control and data flow (static claims) or against an execution trace recorded in an isolated process (dynamic
claims). With step-by-step reasoning, accuracy was saturated: 0.98 on CRUXEval and 1.00 on REval, against 0.88 and
0.89 for direct answers. Of 1{,}257 labelled claims, none of the 337 static claims was wrong; the errors were
confined to run-specific values (3.0\% of 234 dynamic claims on CRUXEval and 0.4\% of 686 on REval), such as
miscounted lengths or loop iterations, and several of them occurred in chains whose final answer was correct. A
verifier that checked the static claims against the program graphs therefore had nothing to correct.

\subsection{Static-Analysis Facts in Vulnerability Audits}
We then gave the model the obligations of our C/C++ analyser (\cref{sec:evidence}) while auditing 30 PrimeVul pairs.
Plain step-by-step auditing judged 10 pairs correctly. Adding the facts gave 11; presenting them as a checklist of
obligations to discharge reduced the number to 5--6, because the model judged more functions SAFE once it had worked
through the listed obligations; a refinement step that asked the model to revisit its verdict in the light of the
concerns gave 9, and five of its seven changed verdicts were patched functions turned into false alarms. Reading the
answers, we found that the model's local claims about the code were almost always correct; its errors were
omissions and judgements, and in several cases it abandoned a correct judgement of its own in favour of a note.

\subsection{Implications for the Main Study}
The two experiments shifted our question. The models do not need static analysis to \emph{see} the structure of the
code; what matters is how they \emph{weigh} what the analysis tells them. This question cannot be answered on 30
pairs or by measuring accuracy alone. The main study therefore manipulates the evidence while keeping the code
fixed, scales the data and the models, and separates exploratory from confirmatory analyses.

\section{Methodology}
\label{sec:method}

\begin{figure*}[t!]
\centering
\includegraphics[width=\textwidth]{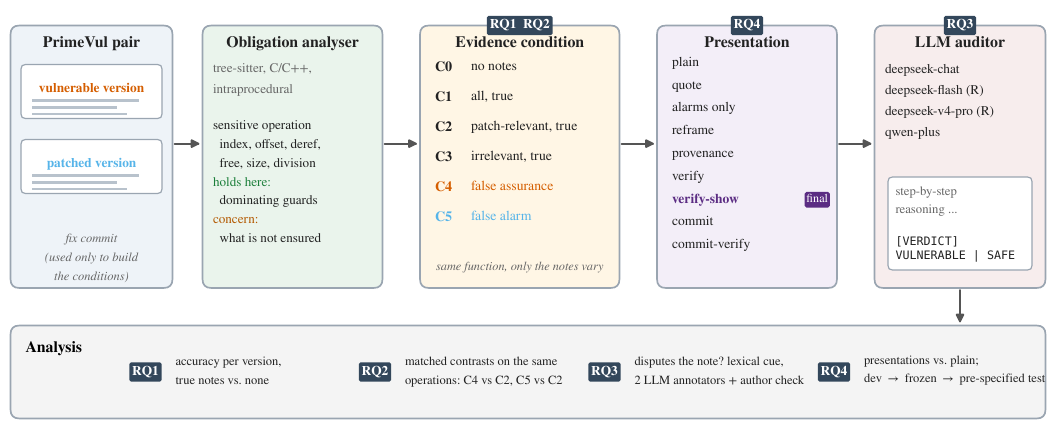}
\caption{Study overview. For each PrimeVul pair, an intraprocedural analyser produces proof obligations for both
versions; the fix commit decides which obligations are relevant and how false evidence is built. The same function is
audited under every evidence condition and, for RQ4, under different presentations of the same notes.}
\label{fig:overview}
\end{figure*}

We study how an LLM that audits a C/C++ function for vulnerabilities uses static-analysis evidence about that
function. The design manipulates the evidence while keeping the code, the model and the task fixed, so that
differences in the model's verdicts can be attributed to the evidence. This section describes the research
questions (\cref{sec:rqs}), the data (\cref{sec:data}), how the evidence is produced and manipulated
(\cref{sec:evidence,sec:conditions}), the measures and statistics (\cref{sec:measures}), and the validation of the
behavioural coding (\cref{sec:annotation}); \cref{sec:setup} describes the models, the audit task and the protocol
that separates exploration from confirmation. \Cref{fig:overview} summarises the design.

\subsection{Research Questions}
\label{sec:rqs}

\textbf{RQ1 (usefulness).} Does realistic static-analysis evidence improve an LLM's vulnerability judgements?

\textbf{RQ2 (relevance and correctness).} How do the relevance and the correctness of the evidence change the
judgements? In particular, do false assurances (``this access is checked'') and false alarms (``this access is
not checked'') affect the model to the same degree?

\textbf{RQ3 (mechanism).} When the evidence is false, does the model check it against the code, and does this
differ between assurances and alarms and across models?

\textbf{RQ4 (presentation).} Can the way evidence is presented reduce deference to false evidence without
losing the value of true evidence?

\subsection{Data}
\label{sec:data}

We use PrimeVul~\cite{ding2024primevul}, a vulnerability dataset with de-duplicated, manually validated
labels. Its \emph{paired} subsets contain, for each vulnerability-fixing commit, the vulnerable version of a
function and its patched version. Pairs are well suited to our questions: a model that truly reasons about
the code should flag the vulnerable version and clear the patched one, and the patch localises the code that
matters. We take the release v0.1 of the paired splits and link the two versions of a pair through their commit.

\paragraph{Study set} From the paired test split (433 linkable pairs) we keep pairs whose CWE (Common Weakness
Enumeration) category describes a structural defect that intraprocedural evidence can speak to (out-of-bounds access:
CWE-119/120/122/125/787; NULL dereference: CWE-476; use after free, double free and leaks: CWE-416/415/401; divide by
zero and integer over/underflow: CWE-369/190/191; infinite loop: CWE-835), whose two versions differ, and whose
versions both have at most 150 lines. This yields 187 pairs (374 functions) from 60 projects; the largest are
TensorFlow (38 pairs), the Linux kernel (26) and Vim (14), and the most frequent CWEs are out-of-bounds write
(CWE-787, 46 pairs), NULL dereference (CWE-476, 32) and out-of-bounds read (CWE-125, 28).

\paragraph{Development and test sets} To evaluate presentation strategies without fitting them to the data
they are evaluated on, we use two further splits that we did not inspect before the corresponding phase
(\cref{sec:protocol}). The development set contains all 237 eligible pairs of the paired validation split. The
test set is a random sample (seed 7) of 300 of the 1{,}390 eligible pairs of the paired training split; since we
do not train models, the training split is simply unseen data. No function hash is shared between the three
splits.

\paragraph{Held-out replication set} An earlier replication (\cref{sec:rq4}) used the 192 pairs of the paired
test split that are disjoint from the study set (any CWE, at most 300 lines).

\subsection{Static-Analysis Evidence: Proof Obligations}
\label{sec:evidence}

To produce evidence whose truth we control, we implemented an intraprocedural analyser for C and C++ on top of
tree-sitter~\cite{treesitter}. For a function it lists the \emph{sensitive operations} (array indexing, pointer
offsets, dereferences of pointers obtained from calls, frees, divisions, copy and allocation sizes, and computed
length arguments) and, for each operation, a \emph{proof obligation} with two parts:
\begin{itemize}
\item \emph{holds here}: the conditions that hold on every path reaching the operation, derived from enclosing
  \texttt{if}/\texttt{else} arms, loop conditions, the left operands of short-circuit operators and ternaries,
  and earlier early-exit checks in enclosing blocks (after \texttt{if (c) return;} the condition \texttt{!c}
  holds);
\item \emph{concern}: what the function itself does not guarantee, for example a missing upper or lower bound
  on an index, a signed index without a non-negativity check, a size computed with a multiplication of
  unbounded values, a use after a free, or a return path that drops an allocation.
\end{itemize}
The analyser handles functions whose return type PrimeVul has stripped from the first line, attribute macros,
\texttt{sizeof} operands, pool allocators and returns that follow a free. It is exact for the syntax it models and
deliberately does not reason about callers and callees; the notes state this. A function's obligations are
rendered as a numbered list, operations with concerns first, at most twelve per function (\cref{fig:notes}).

\begin{figure}[htbp]
\centering
\begin{lstlisting}[basicstyle=\scriptsize\ttfamily,frame=single,columns=fullflexible,keepspaces=true]
O1 [array index] line 13: `avc->pps[pps_id]`
   holds here: `!(pps_id > 255)`
               (early exit, line 11)
   concern: `pps_id` is signed (`s32`) and no
            lower-bound (>= 0) check dominates
            this access
\end{lstlisting}
\caption{Notes for the vulnerable version of a PrimeVul pair (gpac, CWE-120), wrapped to the column width. In the
patched version the check at line 11 becomes \texttt{(pps\_id<0) || (pps\_id > 255)}, the holds-here line changes
accordingly and the concern disappears.}
\label{fig:notes}
\end{figure}

\subsection{Evidence Conditions}
\label{sec:conditions}

For every pair we derive six evidence conditions (\cref{tab:conditions}). Relevance is defined by the patch: an
obligation is \emph{patch-relevant} if its operation, or a line its concerns refer to, lies near a changed hunk
(three lines before to eight lines after) and it mentions an identifier that occurs in the changed lines. The patch
is used only to build the conditions; the model never sees it.

False evidence is constructed so that it can be refuted by reading the function. In \cond{C4} every concern of the
patch-relevant obligations of the vulnerable version is replaced by a fabricated guarantee that cites a concrete
line (for instance ``\texttt{pps\_id >= 0} is checked before this access (line 12)'' where line 12 is
\texttt{return -1;}). In \cond{C5} the patched version receives the concerns that the patch removed, transplanted
onto the same kind of operation, with the contradicting guard facts deleted; when no such concern exists, a true
guard of a patch-relevant (or, failing that, any) operation is declared not to protect it. We record which of the
three constructions was used.

\begin{table*}[t!]
\centering
\caption{Evidence conditions. Functions: number of study-set functions to which a condition applies.}
\label{tab:conditions}
\small
\begin{tabularx}{\textwidth}{@{}llXr@{}}
\toprule
& \textbf{Evidence} & \textbf{Content} & \textbf{Functions} \\
\midrule
\cond{C0} & none & plain audit & 374 \\
\cond{C1} & all, true & every obligation the analyser reports & 374 \\
\cond{C2} & relevant, true & only patch-relevant obligations & 189 \\
\cond{C3} & irrelevant, true & only obligations away from the patch that carry a concern & 183 \\
\cond{C4} & false assurance & vulnerable versions: relevant concerns replaced by fabricated, refutable guarantees & 83 \\
\cond{C5} & false alarm & patched versions: concerns the patch removed, guards deleted & 97 \\
\bottomrule
\end{tabularx}
\end{table*}

\subsection{Measures and Statistics}
\label{sec:measures}

A verdict is correct if it is VULNERABLE for a vulnerable version and SAFE for a patched one. We report accuracy
separately for vulnerable versions (recall) and patched versions (specificity), and the paired outcome of a pair
(both correct, both VULNERABLE, both SAFE, both wrong) following PrimeVul~\cite{ding2024primevul}. Responses
without a verdict count as wrong; API failures are treated as missing.

Following guidelines for experiments in software engineering~\cite{wohlin2012experimentation,arcuri2011practical},
all comparisons are paired on the same functions and we report effect sizes with confidence intervals, not only
$p$-values. We test differences in accuracy with the exact McNemar test~\cite{mcnemar1947note} and report bootstrap
95\% confidence intervals (2{,}000 resamples)~\cite{efron1993bootstrap} of the difference; single proportions have
Wilson intervals. Our main contrasts compare false evidence with true evidence \emph{about the same operations}
(\cond{C4} vs.\ \cond{C2} on vulnerable versions, \cond{C5} vs.\ \cond{C2} on patched versions), which holds the
presence and placement of notes constant and varies only their truth. When several presentation strategies are
compared with a baseline, $p$-values are Holm-corrected~\cite{holm1979simple}.

\paragraph{Direction-aware measure} Accuracy contrasts conflate deference with a model's leaning: a model that
flags almost every function cannot be pushed further towards VULNERABLE by a false alarm, and loses little recall to
a false assurance. We therefore measure how far false evidence moves verdicts towards \emph{its own claim},
relative to true notes about the same operations:
\begin{align}
\Delta_{\mathrm{as}} &= P(\text{SAFE}\mid\cond{C4}) - P(\text{SAFE}\mid\cond{C2}), \label{eq:as}\\
\Delta_{\mathrm{al}} &= P(\text{VULN}\mid\cond{C5}) - P(\text{VULN}\mid\cond{C2}), \label{eq:al}\\
A &= \Delta_{\mathrm{as}} - \Delta_{\mathrm{al}}, \label{eq:asym}
\end{align}
where SAFE and VULN denote a SAFE and a VULNERABLE verdict. The \emph{assurance shift} $\Delta_{\mathrm{as}}$ is computed on vulnerable versions and the \emph{alarm shift}
$\Delta_{\mathrm{al}}$ on patched versions, each on the functions that have both conditions. Each shift lies in
$[-1,1]$: a value of $+1$ means that the false evidence turns every verdict towards its claim, $0$ that it has the
same effect as true notes about the same operations, and negative values that it pushes verdicts away from its claim
(it backfires). The \emph{asymmetry} $A$ lies in $[-2,2]$; it is positive when false assurances move verdicts
towards their claim more than false alarms do, and zero when the two are followed equally, as a pure change in a
model's willingness to flag would produce. No smoothing or guard term is involved. Responses without a verdict count
as neither SAFE nor VULNERABLE. The measure does not capture whether the model's reasoning is correct, only where its
verdicts move. When and why the measure was introduced is described in \cref{sec:protocol}.

\subsection{Validating the Behavioural Coding}
\label{sec:annotation}

For RQ3 we need to know whether an answer disputes a note. LLMs can match or exceed crowd workers on some annotation
tasks~\cite{gilardi2023chatgpt}, but their reliability for software engineering artefacts varies by task and must be
checked against human labels~\cite{ahmed2025annotation}; we therefore validate both the automatic rule and the LLM
annotators. A lexical rule flags answers that state that a note is wrong, inaccurate, a false positive or does not
match the code. To validate it, we drew 80 \cond{C4}/\cond{C5} answers stratified by the rule (40 flagged, 40 not).
Two LLM annotators (Claude and \mdl{deepseek-v4-pro}) labelled each answer independently with a written guide
(Appendix~\ref{app:annotation}): whether it checks a note against the code, whether it rejects a false note, and
whether its verdict relies on a note at face value. Agreement is measured with Cohen's
$\kappa$~\cite{cohen1960kappa}. The first author independently labelled a random subset of 20 answers. Population
rates are estimated by reweighting the strata with the rule's hit rate among all \cond{C4}/\cond{C5} answers.

\section{Experimental Setup}
\label{sec:setup}
This section describes the model configurations compared, the audit task, and the protocol that separates
exploration from confirmation.

\subsection{Models Compared}
\label{sec:models}

We use four model configurations from two vendors: DeepSeek's flash model without reasoning
(\mdl{deepseek-chat}) and with reasoning (\mdl{deepseek-flash}), the larger \mdl{deepseek-v4-pro} with
reasoning, and Alibaba's \mdl{qwen-plus-2025-12-01} without reasoning. Reasoning models run at the
\texttt{low} reasoning effort; at the default effort their reasoning frequently exhausted a 32k-token budget
before a verdict was produced. All calls use temperature~0; budgets are 8k tokens for non-reasoning and 32k for
reasoning models. Because the larger reasoning model needs about six minutes per call, it was run on a random
60-pair subset of the study set and only on \cond{C0}, \cond{C2}, \cond{C4} and \cond{C5}; \mdl{qwen-plus} was
run on \cond{C2}, \cond{C4} and \cond{C5}.

For the registered multi-vendor replication (H3, \cref{sec:protocol}) we added seven further models from four
vendors, all accessed through Alibaba Cloud Bailian's OpenAI-compatible endpoint with reasoning disabled
(\cref{tab:models}). None of them had been run on any data of this study before the hypothesis was registered. They
were run on the study set under \cond{C2}, \cond{C4} and \cond{C5}, which suffices for the contrasts of RQ2 and
RQ3. In total the study covers eleven model configurations from four vendors. All models are used for inference
only; no model training or fine-tuning is performed.

\begin{table*}[t!]
\centering
\caption{Model configurations. R: reasoning enabled (effort \texttt{low}). Sets: S study, H held-out, D development,
T test. Conditions in brackets were run on a subset only.}
\label{tab:models}
\footnotesize\setlength{\tabcolsep}{3.5pt}
\begin{tabular}{@{}llcll@{}}
\toprule
\textbf{Name in this paper} & \textbf{API model identifier} & \textbf{R} & \textbf{Vendor} & \textbf{Sets (conditions)} \\
\midrule
\texttt{deepseek-chat} & \texttt{deepseek-chat} & & DeepSeek & S (all); H, D (C1, C2, C4, C5); T (C2, C4) \\
\texttt{deepseek-flash} (R) & \texttt{deepseek-flash} & \checkmark & DeepSeek & S (all) \\
\texttt{deepseek-v4-pro} (R) & \texttt{deepseek-v4-pro} & \checkmark & DeepSeek & S, 60 pairs (C0, C2, C4, C5) \\
\texttt{qwen-plus} & \texttt{qwen-plus-2025-12-01} & & Alibaba & S (C2, C4, C5); T (C2, C4) \\
\midrule
\multicolumn{5}{@{}l}{\emph{Registered replication (H3), via Alibaba Cloud Bailian}} \\
\texttt{glm-5.2} & \texttt{glm-5.2} & & Zhipu AI & S (C2, C4, C5) \\
\texttt{kimi-k2.6} & \texttt{kimi-k2.6} & & Moonshot AI & S (C2, C4, C5) \\
\texttt{kimi-k3} & \texttt{kimi-k3} & & Moonshot AI & S (C2, C4, C5) \\
\texttt{qwen3.7-plus} & \texttt{qwen3.7-plus-2026-05-26} & & Alibaba & S (C2, C4, C5) \\
\texttt{qwen3.7-max} & \texttt{qwen3.7-max-2026-06-08} & & Alibaba & S (C2, C4, C5) \\
\texttt{qwen3.7-flash} & \texttt{qwen3.7-flash-2026-07-15} & & Alibaba & S (C2, C4, C5) \\
\texttt{deepseek-v4.1-flash} & \texttt{deepseek-v4.1-flash} & & DeepSeek & S (C2, C4, C5) \\
\bottomrule
\end{tabular}
\end{table*}

\paragraph{Why manipulate the evidence rather than use a production analyser?} Three reasons motivate this choice.
First, only constructed evidence lets us know the truth of every note and compare a false note with a true note about
the same operation; with a production tool, the truth of each warning would have to be established by hand and could
not be varied on the same function. Second, the obligation format states guarantees and concerns explicitly, which
is the information that pipelines pass to the model and the distinction our questions are about. Third, keeping the
code, model and task fixed while only the notes change is what makes the contrasts paired and the effects attributable.
We view evidence from production analysers as a complementary, not competing, direction
(\cref{sec:limitations}).

\subsection{Audit Task}
\label{sec:task}

The audit prompt shows the numbered function, the project and file name and, except in \cond{C0}, the notes with a
one-paragraph legend. It asks the model to reason step by step about untrusted values, the operations they reach
and the checks that guard them, and to end with \texttt{[VERDICT] VULNERABLE} or \texttt{[VERDICT] SAFE}. The
prompts are listed in Appendix~\ref{app:prompts}.

\subsection{Exploration and Confirmation}
\label{sec:protocol}

\begin{figure*}[t!]
\centering
\includegraphics[width=\textwidth]{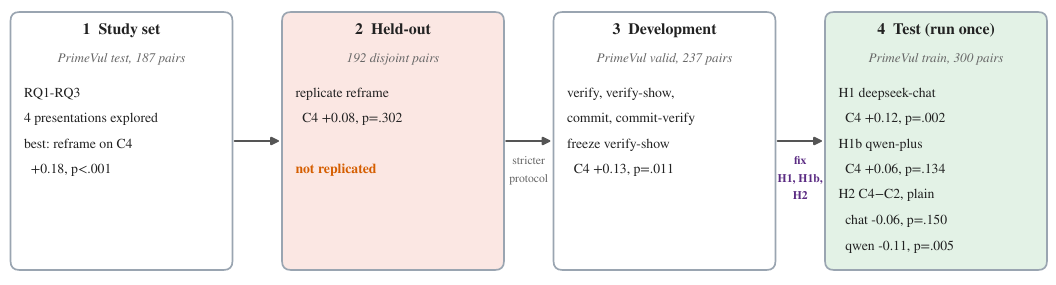}
\caption{Order of the RQ4 evaluations. After the strategy selected on the study set failed to replicate, strategies
were compared on a separate development set only, and the frozen strategy was tested once on unseen pairs with
pre-specified hypotheses. Values are accuracy changes under false assurances relative to the plain presentation
(uncorrected McNemar $p$).}
\label{fig:protocol}
\end{figure*}

RQ1--RQ3 are answered on the study set. For RQ4 we first explored presentation strategies on the study set and
tested the most promising one on the held-out replication set; it did not replicate (\cref{sec:rq4}). We therefore
adopted a stricter protocol (\cref{fig:protocol}): every further strategy was developed and compared on the
development set only; the best one was frozen and the following hypotheses and analyses were fixed before the test set was
run once. Unlike H3 below, they were not written into a separate registration document with hashes, so we call this
test pre-specified rather than registered:
\begin{description}
\item[H1] With the frozen strategy, the accuracy of \mdl{deepseek-chat} on vulnerable versions under false
  assurances (\cond{C4}) is higher than with the plain presentation (paired McNemar test, two-sided,
  $\alpha=0.05$, no further correction).
\item[H1b] The same holds for \mdl{qwen-plus}.
\item[H2] With the plain presentation, accuracy under \cond{C4} is lower than under true relevant evidence
  (\cond{C2}) on the same vulnerable functions.
\end{description}
As a constraint we report the change in accuracy under true relevant evidence (\cond{C2}).

\paragraph{Direction-aware measure and H3} The direction-aware measure (\cref{eq:as,eq:al,eq:asym}) was defined
after the \mdl{qwen-plus} study-set run, that is, after seeing the data of the first four configurations; those
results are therefore exploratory. Before running any further model, we wrote a registration document and recorded
the SHA-256 hashes of it and of the analysis script:
\begin{description}
\item[H3] Over seven new models, the mean asymmetry is positive: false assurances move verdicts towards their claim
  more than false alarms do (bootstrap over pairs, resampled jointly for all models, 2{,}000 resamples, one-sided,
  $\alpha=0.05$).
\end{description}
The registration fixed the models, data, conditions, analysis script (with its hash) and the handling of missing
data (Appendix~\ref{app:h3}). The registration is internal: the document and the hashes were written before the
data were collected, but they were first deposited publicly, with the replication package, after data collection, and
no third party time-stamped the earlier date. We also tested, and rejected before registration, a different post-hoc explanation
(that models follow evidence congruent with their leaning): flip rates conditioned on the baseline verdict were
dominated by regression to the mean, as content-neutral changes of the prompt produced similar flip rates.

\section{Results and Analysis}
\label{sec:results}

We first describe how true evidence affects the audits (RQ1), then contrast true and false evidence about the same
operations (RQ2), examine whether the models check the evidence (RQ3), and finally report every presentation
strategy we evaluated, including the ones that failed (RQ4). Unless stated otherwise, numbers refer to the study set
(187 pairs) and differences are paired on the same functions.

\subsection{RQ1: Does True Static-Analysis Evidence Help?}
\label{sec:rq1}

\begin{figure*}[t!]
\centering
\includegraphics[width=0.85\textwidth]{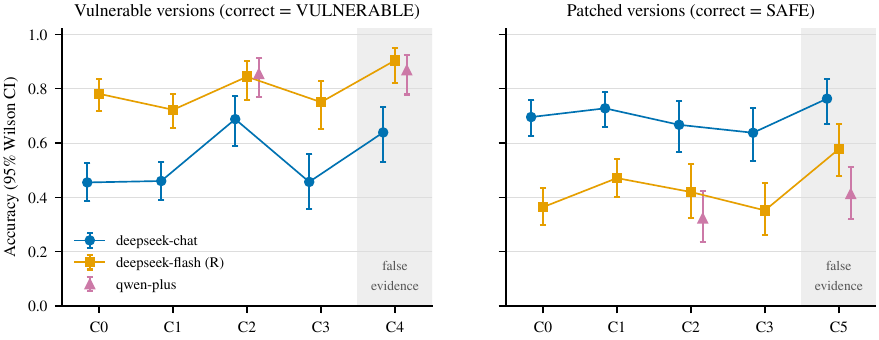}
\caption{Accuracy per evidence condition on the study set, for vulnerable versions (left) and patched versions
(right). Conditions apply to different subsets of functions (\cref{tab:conditions}), so the figure is descriptive;
paired comparisons are in \cref{fig:asymmetry}. False evidence exists only for one version: false assurances
(\cond{C4}) for vulnerable versions and false alarms (\cond{C5}) for patched ones.}
\label{fig:conditions}
\end{figure*}

\Cref{fig:conditions} shows accuracy per condition, separately for vulnerable and patched versions. Two observations
stand out. First, the models differ mainly in their \emph{bias}, not in their discrimination. Without evidence
(\cond{C0}), \mdl{deepseek-chat} judges 36\% of the functions vulnerable and finds 45\% of the vulnerable versions,
whereas \mdl{deepseek-flash} with reasoning judges 69\% vulnerable, finds 78\% of the vulnerable versions, but
clears only 36\% of the patched ones. The larger \mdl{deepseek-v4-pro} is even more inclined to flag (79\% on its
60-pair subset). Reasoning thus shifts the operating point towards suspicion rather than separating the two versions
of a pair better.

Second, true evidence barely moves accuracy. Showing every obligation the analyser reports (\cond{C1}) changes
accuracy by $+0.02$ for both DeepSeek flash configurations (95\% CI $-0.03$ to $+0.07$ and $-0.02$ to $+0.07$;
$p=0.52$ and $p=0.41$). Restricting the notes to the operations that the fix touches (\cond{C2}) yields a larger but
non-significant gain ($+0.06$, CI $-0.01$ to $+0.12$, for \mdl{deepseek-chat}; $+0.04$ for
\mdl{deepseek-flash}), and true notes about other operations (\cond{C3}) have no effect ($+0.00$ and $+0.01$).
For \mdl{deepseek-chat} the small gain under \cond{C2} is spread over both versions ($+0.06$ on vulnerable and
$+0.05$ on patched versions, both n.s.); for \mdl{deepseek-flash} relevant notes counteract its bias, raising
accuracy on patched versions by $0.13$ ($p=0.036$) while lowering it on vulnerable ones by $0.04$ (n.s.). Relevant
notes thus seem to help by pointing the model at the operation that matters, not by telling it something it could
not see in the code. (The models of the replication and \mdl{qwen-plus} were run only on the conditions needed
for RQ2 and RQ3, so RQ1 rests on the two DeepSeek flash configurations and, for \cond{C2} vs \cond{C0}, on
\mdl{deepseek-v4-pro}: $+0.04$, CI $-0.06$ to $+0.15$.)

\begin{answerbox}
\textbf{Answer to RQ1.} Realistic, correct static-analysis evidence does not make the audits measurably more
accurate. Even evidence restricted to the operations that the fix touches yields only a small, non-significant
gain, mostly by drawing attention to the right operation.
\end{answerbox}

\subsection{RQ2: Relevance and Correctness of the Evidence}
\label{sec:rq2}

\begin{figure}[htbp]
\centering
\includegraphics[width=\linewidth]{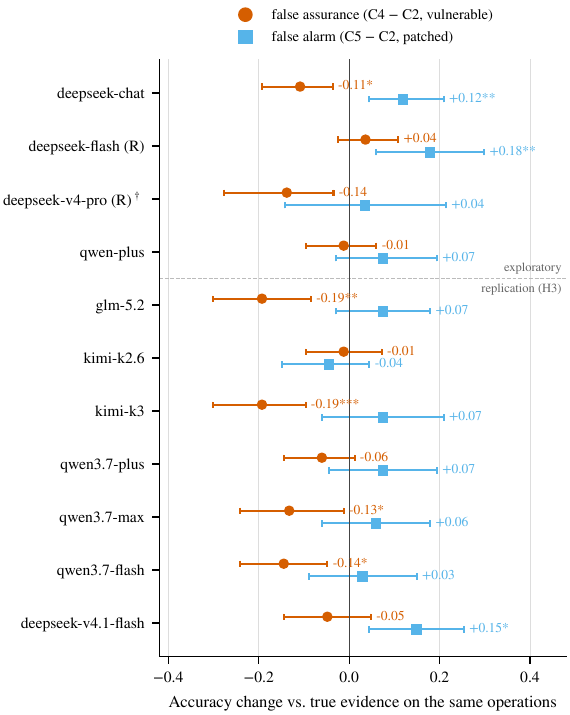}
\caption{Effect of false evidence relative to true evidence about the same operations, with 95\% bootstrap CIs.
Orange: false assurance on vulnerable versions (\cond{C4} $-$ \cond{C2}); blue: false alarm on patched versions
(\cond{C5} $-$ \cond{C2}). The upper block shows the four exploratory configurations and the lower block the seven
models of the registered replication (\cref{sec:h3}). * $p<.05$, ** $p<.01$ (exact McNemar). $^\dagger$60-pair
subset.}
\label{fig:asymmetry}
\end{figure}

To isolate the effect of correctness, we compare false and true notes about the \emph{same} operations of the same
functions (\cref{fig:asymmetry,tab:asymmetry}). The two kinds of false evidence have opposite effects.

\emph{False assurances mislead.} When the notes on a vulnerable function claim that the relevant check exists
(\cond{C4}), the accuracy of \mdl{deepseek-chat} drops from 0.75 to 0.64 relative to the true notes
(\cond{C2}): ten vulnerable functions that were found with true notes are cleared with false ones, and only one moves
in the other direction ($p=0.012$). The larger \mdl{deepseek-v4-pro} shows the same drop on its 60-pair subset
($0.97 \to 0.83$, CI $-0.28$ to $-0.03$, $p=0.125$ given only 29 functions). The reasoning variant of
\mdl{deepseek-flash} and \mdl{qwen-plus} are not misled ($+0.04$ and $-0.01$). Both lean strongly towards
VULNERABLE (\mdl{qwen-plus} flags 77\% of the functions under \cond{C2} and clears only 31\% of the patched
versions), and both find 87--90\% of the vulnerable versions whatever the notes say.

\emph{False alarms do not.} When the notes on a patched function claim that a check is missing (\cond{C5}), accuracy
does not fall; it \emph{rises}, from 0.67 to 0.79 for \mdl{deepseek-chat} (0/8 discordant pairs, $p=0.008$) and
from 0.37 to 0.55 for \mdl{deepseek-flash} ($p=0.008$). A false alarm draws the model to the operation, the model
then finds the guard that the alarm denies, and it clears the function more often than when the true guard is
merely listed. For \mdl{deepseek-v4-pro} ($+0.04$) and \mdl{qwen-plus} ($+0.07$) the change is not
significant, but in no model does a false alarm lower accuracy.

The effect of false evidence therefore depends on its direction. In all four configurations the effect of a false
assurance is below that of a false alarm, and only false assurances ever lower accuracy; how much they do so differs
between models, from no effect for the two models biased towards VULNERABLE to 11--14 points for the others.

\begin{table*}[t!]
\centering
\caption{Matched contrasts on the study set. $n$: functions with both conditions. Discordant: functions correct
only under the true notes / only under the false notes.}
\label{tab:asymmetry}
\footnotesize
\begin{tabular}{@{}llrcccl@{}}
\toprule
\textbf{Model} & \textbf{Contrast} & $\boldsymbol{n}$ & \textbf{True $\rightarrow$ false} & \textbf{Discordant} & $\boldsymbol{p}$ & \textbf{Difference (95\% CI)} \\
\midrule
\mdl{deepseek-chat} & \cond{C4} vs \cond{C2} & 83 & 0.75 $\rightarrow$ 0.64 & 10 / 1 & 0.012 & $-0.11$ ($-0.18$, $-0.04$) \\
 & \cond{C5} vs \cond{C2} & 67 & 0.67 $\rightarrow$ 0.79 & 0 / 8 & 0.008 & $+0.12$ ($+0.04$, $+0.19$) \\
\mdl{deepseek-flash} (R) & \cond{C4} vs \cond{C2} & 83 & 0.87 $\rightarrow$ 0.90 & 3 / 6 & 0.508 & $+0.04$ ($-0.04$, $+0.11$) \\
 & \cond{C5} vs \cond{C2} & 67 & 0.37 $\rightarrow$ 0.55 & 3 / 15 & 0.008 & $+0.18$ ($+0.06$, $+0.30$) \\
\mdl{deepseek-v4-pro} (R)$^\dagger$ & \cond{C4} vs \cond{C2} & 29 & 0.97 $\rightarrow$ 0.83 & 4 / 0 & 0.125 & $-0.14$ ($-0.28$, $-0.03$) \\
 & \cond{C5} vs \cond{C2} & 28 & 0.29 $\rightarrow$ 0.32 & 4 / 3 & 1.000 & $+0.04$ ($-0.14$, $+0.21$) \\
\mdl{qwen-plus} & \cond{C4} vs \cond{C2} & 83 & 0.88 $\rightarrow$ 0.87 & 6 / 5 & 1.000 & $-0.01$ ($-0.08$, $+0.07$) \\
 & \cond{C5} vs \cond{C2} & 67 & 0.31 $\rightarrow$ 0.39 & 5 / 10 & 0.302 & $+0.07$ ($-0.03$, $+0.18$) \\
\bottomrule
\multicolumn{7}{@{}l}{$^\dagger$60-pair random subset of the study set. (R): reasoning enabled.}
\end{tabular}
\end{table*}

\begin{answerbox}
\textbf{Answer to RQ2.} Correctness matters asymmetrically. A fabricated alarm about a fixed function never significantly
lowers accuracy and for two configurations improves it. Relative to true notes about the same operations, a
fabricated assurance that a check exists lowers the detection of real vulnerabilities by 11 points for
\mdl{deepseek-chat} ($p=0.012$); \mdl{deepseek-v4-pro} shows a drop of similar size that is not significant on its
29 functions, and models that already flag most functions are less affected in accuracy terms. \Cref{sec:h3}
examines the asymmetry across vendors with a direction-aware measure. True evidence helps only when it is relevant, and then only
slightly.
\end{answerbox}

\subsection{RQ3: Do the Models Check the Evidence?}
\label{sec:rq3}

\begin{figure*}[t!]
\centering
\includegraphics[width=\textwidth]{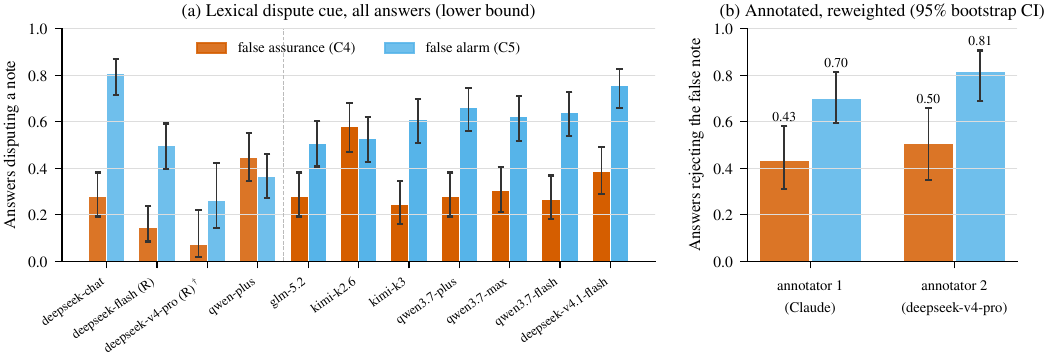}
\caption{(a) Share of answers that dispute a note according to the lexical cue, by model and type of false evidence
(Wilson 95\% CIs). (b) Share of answers that reject the false note according to each annotator, reweighted from the
80-answer stratified sample to all \cond{C4}/\cond{C5} answers (bootstrap 95\% CIs).}
\label{fig:scrutiny}
\end{figure*}

The asymmetry suggests that the models scrutinise alarms but accept assurances. \Cref{fig:scrutiny}(a) supports this
with the lexical cue: \mdl{deepseek-chat} disputes a false alarm in 80\% of its answers (78/97) but a false
assurance in only 28\% (23/83); for \mdl{deepseek-flash} the rates are 49\% and 14\%, and for
\mdl{deepseek-v4-pro} 26\% and 7\%. \mdl{qwen-plus} is the exception: it disputes false assurances
\emph{more} often than false alarms (45\%, 37/83, versus 36\%, 35/97). When it disputes a false assurance, its verdict
is almost always correct (35 of 37), which is consistent with its immunity to false assurances in RQ2; when it
disputes a false alarm, it still judges the patched function vulnerable in 14 of 35 cases. We read a sample of the
flagged \mdl{qwen-plus} answers to make sure that the cue, validated on DeepSeek answers, captures the same
behaviour; it does (for instance, ``line 18 is a comment, not a check. That means the analysis note is incorrect or
misleading''). The cue is conservative: against the annotations its precision is 0.97--1.00
and its recall 0.70--0.76, so the absolute rates are lower bounds, but there is no reason for the misses to depend on
the condition.

The annotations confirm the gap (\cref{fig:scrutiny}(b)). The two LLM annotators agreed well on whether an answer
rejects the false note ($\kappa = 0.82$) and on whether its verdict relies on a note at face value ($\kappa=0.75$),
less on whether an answer checks the notes at all ($\kappa=0.52$, raw agreement 85\%). The first author's
independent labels of 20 answers agreed with both annotators on every item. Reweighted to all \cond{C4}/\cond{C5}
answers, the share of answers that reject the false note is 0.43--0.50 for false assurances and 0.70--0.81 for false
alarms, depending on the annotator.

The answers show how the asymmetry arises. Faced with an alarm, the models typically go to the cited operation, quote
the guard and conclude that the note ``appears to be a false positive''. Faced with an assurance, they more often
restate it as a fact (``the static analysis notes confirm \ldots'') and move on to other operations. A claim that
a check is \emph{missing} invites a search for the check; a claim that it is \emph{present} ends the search.

\paragraph{Can the models refute false assurances when asked to?} A different reading is that false assurances are
simply harder to refute: refuting an alarm means finding a check that exists, whereas refuting an assurance means
establishing that a check is absent. To separate not checking from not being able to check, we ran the verification
call of verify-show (Appendix~\ref{app:prompts}), unchanged, with \mdl{deepseek-chat} on every claim of the
\cond{C2}, \cond{C4} and \cond{C5} notes of the study set (357 calls). A claim of \cond{C4} or \cond{C5} counts as false
if it does not occur among the true \cond{C2} claims of the same function. When asked to check, the model refuted
false assurances about as often as false alarms (\cref{tab:verify}): 0.76 versus 0.79 of the claims. It also
discriminated well between true and false assurances (2\% of true holds-here claims refuted), whereas it refuted 45\%
of the true concerns, a general scepticism towards alarms that matches the backfire in RQ2. At the level of functions
the picture is less even: on the 57 pairs with both conditions, all false claims were refuted for 36 vulnerable
versions under \cond{C4} and for 48 patched versions under \cond{C5} (difference $-0.21$, 95\% CI $-0.37$ to
$-0.05$). Among the false alarms, those that negate a guard of an arbitrary operation were refuted less often (0.70)
than transplanted concerns (0.93) and negated guards of relevant operations (1.00). Overall, the model is able to
refute most false assurances once it is asked to, which supports the reading that the asymmetry in the audits arises
mainly from whether the model checks a claim rather than from whether it can; the residual function-level difference
shows that the two cannot be fully separated in this design, since the direction of a claim is tied to whether
refuting it requires finding a check or establishing its absence. This check used one model.

\begin{table}[htbp]
\caption{Verification of true and false claims when the model is asked to check them (\mdl{deepseek-chat}, study
set). Shares of claims; refuted with Wilson 95\% CI.}
\label{tab:verify}
\centering
\footnotesize
\setlength{\tabcolsep}{3pt}
\begin{tabular}{@{}lrccc@{}}
\toprule
\textbf{Claims} & $\boldsymbol{n}$ & \textbf{Refuted} & \textbf{Confirmed} & \textbf{Unclear} \\
\midrule
\cond{C4} false assurances & 251 & 0.76 (0.70, 0.81) & 0.10 & 0.14 \\
\cond{C5} false alarms & 428 & 0.79 (0.75, 0.82) & 0.06 & 0.15 \\
\cond{C2} true holds-here & 203 & 0.02 (0.01, 0.06) & 0.90 & 0.08 \\
\cond{C2} true concerns & 488 & 0.45 (0.41, 0.50) & 0.31 & 0.24 \\
\bottomrule
\end{tabular}
\end{table}

\begin{answerbox}
\textbf{Answer to RQ3.} The three DeepSeek configurations verify false alarms against the code far more often than
false assurances (annotated: roughly three in four versus one in two answers). We call this pattern \emph{asymmetric
scrutiny}: evidence that the code is safe is trusted, evidence that it is unsafe is checked. It is not universal:
\mdl{qwen-plus} checks assurances at least as often as alarms, and it is also the model that false assurances do
not mislead. When explicitly asked to check, \mdl{deepseek-chat} refutes false assurances about as often as false
alarms (0.76 vs.\ 0.79 of the claims), so the gap lies mainly in whether the claims are checked.
\end{answerbox}

\subsection{Replication Across Vendors (H3)}
\label{sec:h3}

\begin{figure*}[t!]
\centering
\includegraphics[width=\textwidth]{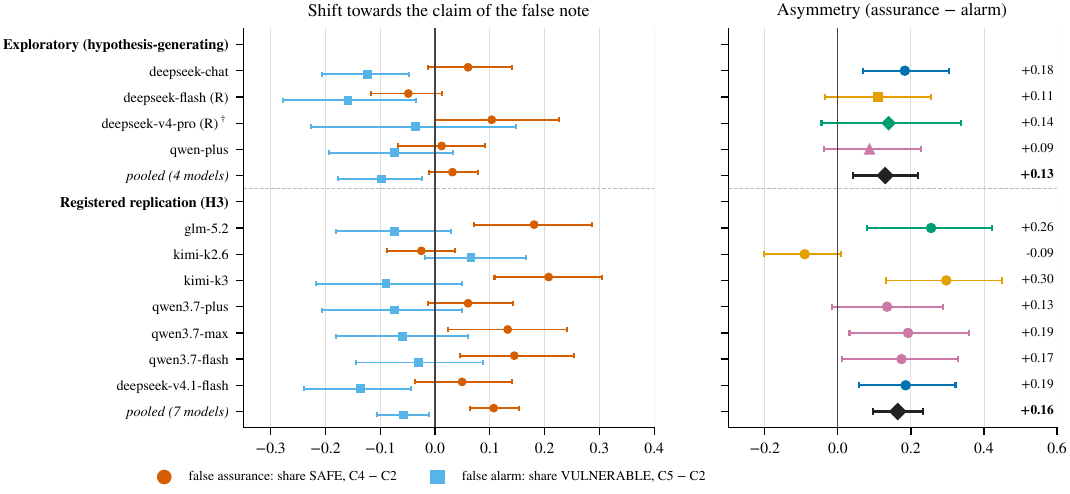}
\caption{How far false evidence moves verdicts towards its own claim, relative to true notes about the same
operations (left), and the asymmetry between false assurances and false alarms (right), with 95\% bootstrap CIs over
pairs. Top: the four configurations from which the measure was derived; bottom: the seven models of the registered
replication. Diamonds: mean over the models of a group. $^\dagger$60-pair subset.}
\label{fig:direction}
\end{figure*}

\Cref{fig:direction} shows the direction-aware measure (\cref{eq:as,eq:al,eq:asym}) for all eleven configurations.
In the four exploratory configurations, false assurances moved verdicts towards SAFE by $+0.03$ on average, while
false alarms moved verdicts \emph{away} from VULNERABLE ($-0.10$); the mean asymmetry was $+0.13$ (95\% CI $+0.04$
to $+0.22$).

The registered replication confirmed H3. Over the seven new models the mean asymmetry was $+0.16$ (95\% CI $+0.10$
to $+0.24$, one-sided bootstrap $p<0.001$). It was positive for six of the seven models (sign test, secondary,
$p=0.06$) and its confidence interval excluded zero for five. False assurances shifted the new models towards SAFE by
$+0.11$ on average, and false alarms shifted them away from VULNERABLE by $-0.06$: the backfire of false alarms that
we observed for DeepSeek also appears in other vendors' models. The exception is \mdl{kimi-k2.6}, which leans
strongly towards VULNERABLE (79\% under \cond{C2}, 22\% accuracy on patched versions), is moved by neither kind of
false evidence, and gives no verdict in 24 of its 369 answers.

In accuracy terms (\cref{fig:asymmetry}), false assurances lowered the detection of vulnerable versions significantly
for four of the seven new models: \mdl{glm-5.2} and \mdl{kimi-k3} by 0.19 ($p=0.002$ and $p=0.001$),
\mdl{qwen3.7-flash} by 0.14 ($p=0.012$) and \mdl{qwen3.7-max} by 0.13 ($p=0.035$). No false alarm lowered
accuracy significantly in any of the eleven configurations. The lexical cue repeats the pattern of RQ3: six of the
seven new models disputed false alarms more often than false assurances (\cref{fig:scrutiny}(a)), for instance
\mdl{kimi-k3} 61\% versus 24\% and \mdl{deepseek-v4.1-flash} 75\% versus 39\%; \mdl{kimi-k2.6} disputed
both at similar rates (53\% and 58\%).

\begin{answerbox}
\textbf{Replication.} Across eleven models from four vendors, false assurances pull verdicts towards their claim,
while false alarms are checked and often backfire. The asymmetry was registered and confirmed on seven models not
used to derive it ($+0.16$, 95\% CI $+0.10$ to $+0.24$); its size varies between models, and a model that flags
nearly every function is moved by neither.
\end{answerbox}

\subsection{RQ4: Presenting Evidence}
\label{sec:rq4}

\begin{figure*}[t!]
\centering
\includegraphics[width=\textwidth]{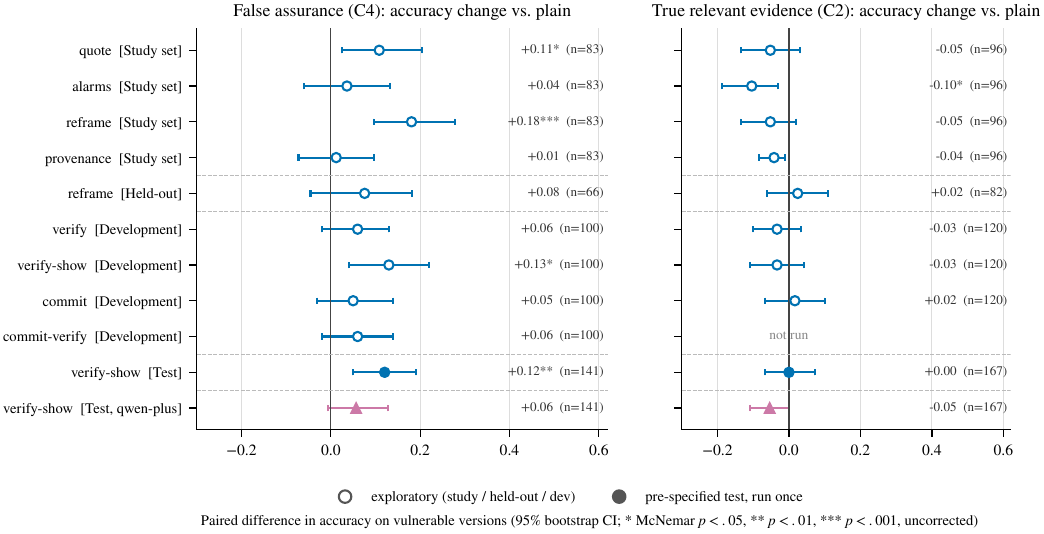}
\caption{Every presentation strategy evaluated, in the order of evaluation: accuracy change relative to the plain
presentation on vulnerable versions under false assurances (left) and under true relevant evidence (right). Hollow
markers: exploratory comparisons; filled markers: the pre-specified test, run once. Blue: \mdl{deepseek-chat};
pink: \mdl{qwen-plus}.}
\label{fig:design_space}
\end{figure*}

Since the models already check alarms, we looked for presentations that make them check assurances too, without
losing what true evidence offers. \Cref{tab:presentations} lists the strategies along the dimensions of the design
space: what is shown, how it is framed, whether its source is shown, and when and by whom it is checked.
\Cref{fig:design_space} reports every strategy we evaluated, in the order in which we evaluated them
(\cref{fig:protocol}).

\begin{table*}[t!]
\centering
\caption{Presentation strategies. All use the same notes; only their presentation changes.}
\label{tab:presentations}
\small
\begin{tabularx}{\textwidth}{@{}llX@{}}
\toprule
\textbf{Strategy} & \textbf{Dimension} & \textbf{Change relative to \emph{plain}} \\
\midrule
quote & framing & instruction to treat notes as unverified claims and to quote the code before relying on one \\
alarms & content & ``holds here'' assurances removed; operations and concerns kept \\
reframe & framing & assurances recast as claims to verify (``to verify: the tool reports \ldots'') \\
provenance & source & the source text of every line a note cites is printed next to the citation \\
verify & checking & a separate call checks each claim against the code; refuted claims are withheld \\
verify-show & checking & as verify, but refuted claims are listed separately as contradicted by the code \\
commit & timing & the model first audits without notes, then reconsiders with them (two turns) \\
commit-verify & timing + checking & commit with the notes of verify-show \\
\bottomrule
\end{tabularx}
\end{table*}

\paragraph{Exploration on the study set} Of the first four strategies, reframing assurances as claims to verify was
by far the most effective: accuracy under false assurances rose from 0.64 to 0.82 (16 functions gained, one lost;
Holm-adjusted $p=0.001$) with no significant loss under true evidence. Asking the model to quote the code before
relying on a note helped less ($+0.11$, Holm $p=0.15$). Removing assurances altogether (alarms) and showing provenance
had no effect on false assurances, and removing assurances cost 10 points of recall under true relevant evidence
($p=0.021$): the assurances that mislead under \cond{C4} are useful under \cond{C2}. Showing provenance did not help
even though the printed source line contradicted the false note; the models did not use the contradiction unless
asked to check.

\paragraph{A failed replication} Reframing was selected on the same functions on which it was evaluated. To check
that the gain was not an artefact of this selection, we pre-specified a replication on the 192 held-out pairs. It did
not replicate: accuracy under false assurances rose from 0.38 to 0.45 (10 gained, 5 lost, $p=0.30$). Deference itself
was also weaker on these pairs (gap $-0.09$, $p=0.18$), which leaves less to recover, but the gain was clearly smaller
than on the study set. We take this as evidence that the selection had overestimated the effect, and we changed the
protocol: all further strategies were developed on a separate development set, and one was frozen and tested once.

\begin{figure*}[t!]
\centering
\includegraphics[width=\textwidth]{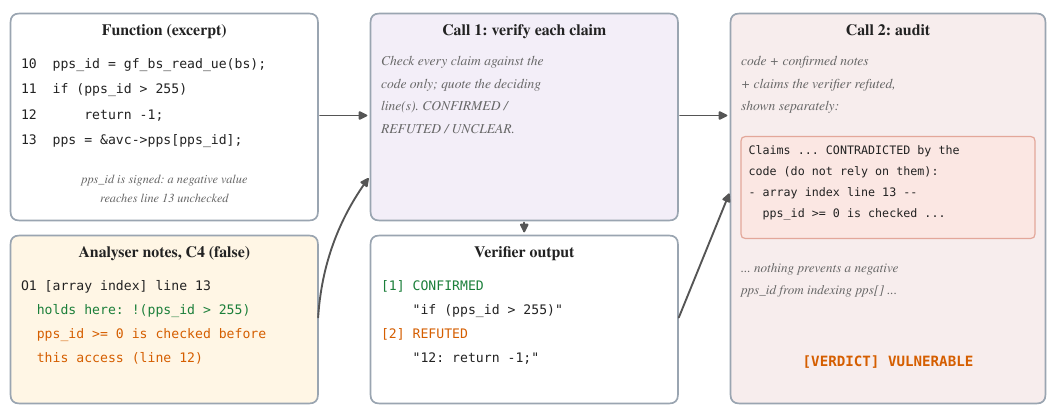}
\caption{Verify-show on a false assurance (illustrative, simplified from a study-set pair). A first call checks each
claim of the notes against the code and quotes the deciding line; the audit then sees the confirmed notes and,
separately, the claims that the code contradicts.}
\label{fig:verify_show}
\end{figure*}

\paragraph{Development} On the development set we compared four strategies that move the checking out of the audit
or before it (\cref{fig:design_space}, development rows). Checking each claim in a separate call and listing the
refuted claims next to the confirmed notes (verify-show, \cref{fig:verify_show}) raised accuracy under false
assurances from 0.59 to 0.72 (18 gained, 5 lost; Holm-adjusted $p=0.032$) without a significant loss under true
evidence. Withholding the refuted claims instead (verify) helped less ($+0.06$): the explicit statement that the code
contradicts a note appears to matter, not just its absence. Committing to an initial verdict before seeing the notes
(commit) was the only strategy that made \emph{true} evidence useful ($+0.05$ under \cond{C1}, $p=0.023$; $+0.09$ on
patched versions under \cond{C2}, $p=0.076$), but it did little against false assurances ($+0.05$), and combining it
with verification did not add the two effects ($+0.06$ under \cond{C4}). Following the rule we had fixed in advance,
verify-show became the final strategy.

\paragraph{Pre-specified test} On the 300 test pairs, run once after the hypotheses were fixed, verify-show raised
the accuracy of \mdl{deepseek-chat} under false assurances from 0.54 to 0.66 ($+0.12$, CI $+0.05$ to $+0.19$; 23
functions gained, 6 lost; $p=0.002$), confirming H1. Accuracy under true relevant evidence was unchanged ($+0.00$,
CI $-0.04$ to $+0.05$, on 338 functions). The share of \cond{C4} answers that dispute a note doubled, from 56 to 114
of 141.

H1b, the same hypothesis for \mdl{qwen-plus}, was not confirmed. Verify-show raised its accuracy under false
assurances from 0.78 to 0.84 ($+0.06$, CI $-0.01$ to $+0.13$; 15 gained, 7 lost; $p=0.13$), in the same direction
as for \mdl{deepseek-chat} but half as large, and accuracy on vulnerable versions under true relevant evidence fell
by 0.05 (CI $-0.11$ to $+0.00$, $p=0.11$). Verification also changed the answers of \mdl{qwen-plus} less: the share
of \cond{C4} answers that dispute a note did not increase (61 vs.\ 54 of 141).

H2, the replication of deference, gave mixed results. For \mdl{deepseek-chat}, accuracy under false assurances was
0.06 lower than under true evidence on the same functions, but the difference was not significant (CI $-0.14$ to
$+0.01$, $p=0.15$); the direction agrees with the study set ($-0.11$) and the held-out set ($-0.09$). For
\mdl{qwen-plus}, which false assurances had not misled on the study set, the drop on the test set was 0.11 (from
0.89 to 0.78, CI $-0.19$ to $-0.04$, $p=0.005$). Deference to false assurances is therefore not a property of one
model, but its size varies between models and samples. In an analysis that we did not pre-specify and therefore label
post hoc, we pooled the disjoint sets on which each model was run under the plain presentation. For
\mdl{deepseek-chat} (study, held-out and test sets; 290 vulnerable functions) false assurances lowered accuracy by
0.08 (from 0.61 to 0.53, CI $-0.13$ to $-0.03$; 40 lost, 16 gained; $p=0.002$); for \mdl{qwen-plus} (study and test
sets; 224 functions) by 0.08 as well (from 0.89 to 0.81, CI $-0.13$ to $-0.02$; 29 lost, 12 gained; $p=0.012$).

\begin{answerbox}
\textbf{Answer to RQ4.} Instructions and framing inside the audit prompt did not reliably reduce deference: the one
that worked on the study set did not replicate. Moving verification into a separate step and showing its refutations
did: in a pre-specified test it recovered 12 points of accuracy under false assurances for \mdl{deepseek-chat} at no
cost under true evidence. For \mdl{qwen-plus} the gain was smaller ($+0.06$) and not significant.
\end{answerbox}

\section{Discussion}
\label{sec:discussion}

\subsection{Why Are Assurances Trusted and Alarms Checked?}

The answers suggest a simple mechanism. An alarm states that something is missing at a specific operation; the
natural response is to look for it, and when the model looks, it usually finds the guard that the alarm denies. An
assurance states that the work has already been done; it gives the model a reason to stop looking at that operation
and to spend its reasoning elsewhere. In this sense the asymmetry is not a failure to read the code (the models read
the relevant lines correctly when they look at them, \cref{sec:preliminary}) but a failure to look. The same
allocation of attention is familiar from human automation bias, where errors of commission arise when an aid's
output replaces the verification it was meant to
support~\cite{parasuraman1997humans,skitka1999automation,goddard2012automation}. In the terms of Lee and
See~\cite{lee2004trust}, the models' reliance on the tool is miscalibrated in one direction only. We read the answers
as consistent with this mechanism; the study does not demonstrate it directly.

Other explanations are compatible with the data. Assistants trained to agree with their users may agree more
readily with statements that reduce the scope of the task~\cite{sharma2024sycophancy}, and the security-oriented
framing of our prompts may make alarms feel more salient than reassurances. Our study does not separate these
explanations; it establishes that the asymmetry exists, that it is not specific to one model or vendor, and that it
also appears in a direction-aware measure that compares each kind of false evidence with true notes about the same
operations. It also qualifies the conclusion of concurrent work that
contextual cues mainly change a model's willingness to flag code~\cite{shahriar2026words}: for claims that can be
checked against the code, the direction of the claim decides whether it is checked. The observation that blunt
attempts to bias LLM code reviewers can raise suspicion~\cite{alexopoulos2026contextual} fits the same pattern:
the models scrutinise what looks like an attempt to change their verdict, and an alarm is such an attempt in a way
that a quiet assurance is not.

\subsection{Failure Modes and Boundary Conditions}

Two boundary conditions recur in the data. First, a model's leaning limits how much any evidence can move it: the
exception \mdl{kimi-k2.6} flags almost every function, has little to gain from scrutiny and little to lose to an
assurance, and is moved by neither kind of false evidence. Second, the size of deference varies between samples of
the same model. On the study set \mdl{qwen-plus} appeared immune to false assurances because it flags most functions;
on the test set the same model lost 11 points to them, and the direction-aware measure showed the same tendency as
for the other models. Accuracy is therefore a poor measure of deference. The models we studied differ more in their
leaning than in their discrimination (\cref{sec:rq1}), and a model's leaning determines how much room evidence has to
move it in either direction. Studies that evaluate LLM pipelines with analysis evidence should report, at least, the
shift towards the evidence's claim relative to a matched baseline, separately for evidence that argues for and
against a vulnerability. The mitigation has its own boundary: verify-show helped \mdl{deepseek-chat} in the
pre-specified test but gave a smaller, non-significant gain for \mdl{qwen-plus}, whose answers it changed less.

\subsection{Implications for LLM--Analysis Pipelines}
\label{sec:implications}

\paragraph{Treat the tool's guarantees as the riskiest input} Pipelines that let an LLM triage static-analysis alarms
rely on the direction of evidence that the models already check. Pipelines that pass the analyser's guarantees to the
model (for example that a value is bounded, that a pointer is checked, or that a path is infeasible) rely on the
direction that the models trust. An incorrect guarantee, whether caused by an unsound analysis, a stale result or an
adversary, is likely to propagate into the model's verdict. Guarantees should be verified before they reach the
auditor, or presented with their provenance and an explicit instruction to verify them in a separate step.

\paragraph{Separate verification from judgement} Instructions inside the audit prompt were fragile: asking the model
to quote the code before relying on a note helped little, and reframing assurances as claims to verify, the best
instruction on the study set, did not replicate. Showing the source line that contradicts a note did not help either;
the models did not use the contradiction unless asked to check. What worked was moving the check out of the audit:
a separate call verifies each claim against the code, and the audit receives the result. This mirrors
Chain-of-Verification, where answering verification questions independently of the draft is essential to its
effect~\cite{dhuliawala2024cove}, and is consistent with the finding that self-correction succeeds mainly when it can
draw on a reliable external reference~\cite{kamoi2024selfcorrection}, here the code itself. It also matches the
human finding that explanations reduce over-reliance when they make verification
easy~\cite{vasconcelos2023explanations}, and the adversarial finding that isolating the code from unverifiable
comments defends LLM detectors better than instructions to distrust them~\cite{wu2026alibi}. Verify-show doubles the
number of calls; the verification call is short, since it returns a JSON array of statuses, but its cost is not
negligible. Withholding refuted claims (verify) was less effective than showing them as contradicted: an explicit
statement that the code contradicts a claim appears to correct the auditor, whereas the mere absence of a claim does
not.

\paragraph{Make true evidence useful} True evidence did little in our study, and only the commit-first strategy, in
which the model audits the function before seeing the notes and then reconsiders, made it measurably useful on the
development set. Commit-first is a cognitive forcing function in the sense of Bu\c{c}inca et
al.~\cite{bucinca2021trust}: it ensures that the model forms its own judgement before the tool can anchor it. Its
effects did not add up with those of verification in our combination, and it was not tested on the test set; whether
a better combination exists is an open question.

\paragraph{Implications for research} The failed replication of the reframing strategy is a cautionary example.
Selected among four strategies on 83 functions, it raised accuracy by 18 points with a Holm-adjusted $p$ of 0.001,
yet on 66 held-out functions the gain fell to 8 points and was not significant. Prompt variants are cheap to generate
and to evaluate, which makes selection effects of this kind easy to produce and hard to notice. Separating
development and test data, freezing the strategy, and fixing the hypothesis before the test is run cost us one
additional dataset and a few days; it is what allowed us to report the verify-show result with confidence, and the
H1b result with honesty.

\subsection{Limitations and Threats to Validity}
\label{sec:limitations}

\paragraph{Construct validity} Our evidence is produced by our own analyser and manipulated by construction, not
taken from a production tool. This is deliberate, since it lets us control the truth and relevance of every note, but
real tools phrase their output differently and are wrong in different ways. Our false assurances are refutable by
reading the function, and they cite a concrete line; real incorrect guarantees may be harder to refute, which would
make the deference we measure a lower bound. Relevance is defined through the fix, which approximates but does not
equal what matters for the vulnerability. The behavioural coding of RQ3 relies on a lexical cue; we validated it
against two independent LLM annotators and a human check of 20 items, found it conservative (precision 0.97--1.00,
recall 0.70--0.76), and report reweighted annotated rates. The annotators are LLMs; the agreement with the first
author's labels was perfect on the checked subset, but the subset is small. The cue was validated on DeepSeek
answers; we read a sample of flagged \mdl{qwen-plus} answers but did not formally validate it for the models of
the replication.

\paragraph{Internal validity} All comparisons are paired on the same functions and differ only in the notes or their
presentation. Responses at temperature~0 are not fully deterministic~\cite{ouyang2025nondeterminism}, so some
discordant pairs reflect noise; noise affects both directions of a McNemar test equally and cannot create a
systematic difference, but it lowers power. The direction-aware measure was defined after seeing the first four
configurations; we therefore treat those results as exploratory and confirmed the measure on seven new models under
an internal registration (\cref{sec:protocol}). We also examined, and rejected before the registration, an alternative analysis (flip rates
conditioned on the baseline verdict) after finding that it is dominated by regression to the mean. Responses without
a verdict are counted as wrong in accuracy and as neither verdict in the direction-aware measure, which is why the
two can differ (for \mdl{deepseek-chat}, two of the ten vulnerable functions lost under \cond{C4} were left without
a verdict rather than judged SAFE); they are rare (at
most 3.2\% for all configurations except \mdl{kimi-k2.6} with 6.5\%).

\paragraph{External validity} We study C/C++ functions of at most 150 lines (300 on the held-out set) with
structural CWEs, audited in isolation; longer functions, other languages, inter-procedural evidence and agentic
settings may behave differently. PrimeVul's labels are validated but not free of noise, and the vulnerable version of
a pair may contain issues beyond the one fixed. The models are accessed through vendor APIs; we pin dated snapshots
where the vendor offers them, but undated model names may change behind the API, and we cannot rule out that
PrimeVul functions appear in the models' training data~\cite{sallou2024breaking}. Such contamination would affect
all conditions of a function alike and cannot explain differences between them. Results may also depend on the
wording of our prompts~\cite{sclar2024quantifying}; we used one wording per condition and presentation, and the
replication across eleven configurations reduces, but does not eliminate, this threat. Models were run without
reasoning, except for two DeepSeek configurations at low reasoning effort; the reasoning configurations showed the
same asymmetry, but we did not vary the reasoning effort. Results for \mdl{deepseek-v4-pro} are based on a 60-pair
subset. The mitigation was tested on two models; its effect on \mdl{qwen-plus} was half as large and not
significant, so we do not claim that it generalises across models.

\paragraph{Conclusion validity} We report paired exact tests, bootstrap confidence intervals and Holm corrections
where several strategies are compared with a baseline. The confirmatory tests use the analyses fixed in advance; only H3 has a registration document with hashes, and that
registration is internal (\cref{sec:protocol}). The pooled analysis of deference across the three sets
(\cref{sec:rq4}) was not pre-specified and is labelled post hoc.
Several per-model results rest on 60--90 functions and have wide intervals; we base our conclusions on the pooled and
confirmatory tests rather than on individual models.

\subsection{Reproducibility of the Analysis Scripts}

All code used in this study is released as a companion repository at
\url{https://github.com/lzh0082-cloud/CODE-DETECTION} under the MIT license. It contains the obligation analyser,
the scripts that build the evidence conditions and presentations, run the audits and the verification calls, and
compute every table and figure in this paper, together with the annotations and the registrations with their hashes.
The full run records, with the evidence of all conditions and all model responses, and the cache of all API responses
are archived separately and linked from the repository. PrimeVul is a public dataset and is not redistributed; the
repository explains how to obtain its paired splits from the original release.

\section{Conclusion}
\label{sec:conclusion}
We studied how LLM vulnerability auditors use static-analysis evidence by auditing the same functions under
evidence that differed only in relevance, truth and presentation, on 916 PrimeVul pairs with eleven model
configurations from four vendors.

\begin{itemize}
\item \textbf{True evidence barely improved the audits.} Showing all true notes changed accuracy by $+0.02$; relevant
  notes gave a small, non-significant gain.
\item \textbf{False evidence acted asymmetrically.} Fabricated assurances that a check exists pulled verdicts towards
  SAFE and lowered the detection of real vulnerabilities, whereas fabricated alarms were checked against the code and
  often made the model clear the function. The asymmetry was confirmed in an internally registered replication on seven models
  that were not used to derive it; its size varies between models and samples.
\item \textbf{Separate verification helped; instructions did not reliably.} Instructions within the audit prompt did
  not reliably reduce deference, and the most promising one failed to replicate; verifying each claim in a separate
  step and showing the refuted claims did, in a pre-specified test for one of two models, without reducing the value of
  true evidence.
\end{itemize}

\paragraph{Practical implications}
\begin{itemize}
\item Treat the analyser's guarantees, not its alarms, as the input that most needs checking in LLM--analysis
  pipelines.
\item Check guarantees outside the auditing prompt, in a separate verification step whose refutations are shown to
  the auditor.
\item Measure deference in a direction-aware way, and separate the development of prompts from their evaluation.
\end{itemize}

\paragraph{Limitations and future work} The findings are bounded by C/C++ functions audited in isolation, evidence
from our own analyser, and models accessed through vendor APIs. Future work should (i) study evidence from production
analysers; (ii) extend the study to inter-procedural and agentic settings; (iii) test verification strategies on more
models; and (iv) find ways to make true evidence useful without making false evidence harmful.

\section*{Acknowledgment}
This work was supported by aSSIST University.

\section*{Ethics and Data Use}
The study uses only PrimeVul, a public dataset of open-source code, and the outputs of commercial LLM APIs. PrimeVul is used under the terms of its release and is not redistributed.

\section*{Disclosure on the Use of LLMs in Manuscript Preparation}
LLMs are the object of this study; the models described in the Experimental Setup were the subject of study and were
not used for manuscript writing. 
All scientific claims, experimental
designs, data analyses, and interpretations are the authors’
own, and the authors take full responsibility for the content
of the paper.
In addition, two LLMs (Claude and \mdl{deepseek-v4-pro}) served as annotators for
the validation of the behavioural coding, as described in \cref{sec:annotation}. 

\appendix

\lstset{basicstyle=\scriptsize\ttfamily,breaklines=true,frame=single,columns=fullflexible,keepspaces=true}

\section{Prompts}
\label{app:prompts}

All prompts are reproduced verbatim from the replication package. \verb|{...}| marks inserted content.

\subsection{Audit}

\paragraph{System message} It reads:
\begin{lstlisting}
You are an expert C/C++ security auditor who reasons precisely about code.
\end{lstlisting}

\paragraph{User message} The function is shown with line numbers. Under \cond{C0} the notes block is omitted.
\begin{lstlisting}
Decide whether the following function (project `{project}`, file `{file}`) contains a security vulnerability. Line numbers are shown for reference only.

```c
{numbered function}
```

Static analysis notes for reference. {legend}

{notes}

Reason step by step, citing line numbers: identify untrusted or externally controlled values, trace how they flow into memory accesses, pointer dereferences, frees, arithmetic and calls, and state which checks guard each of them. Then finish with a final line:
[VERDICT] VULNERABLE  or  [VERDICT] SAFE
\end{lstlisting}

\paragraph{Legend (plain presentation)} It reads:
\begin{lstlisting}
'holds here' lists conditions guaranteed on every path that reaches the operation; 'concern' states what this function itself does not guarantee. The analysis is exact for this function's code but does not look into callers or called functions.
\end{lstlisting}

\subsection{Presentation Strategies}

\paragraph{quote} The following paragraph is inserted between the notes and the task:
\begin{lstlisting}
Treat every note as an unverified claim, not as a fact. Before you rely on any 'holds here' or 'concern' line, quote the exact code it refers to and say whether that code confirms the claim. Do not rely on a note that you could not confirm in the code; reason from the code instead.
\end{lstlisting}

\paragraph{alarms} All \texttt{holds here} lines are removed, and the legend becomes:
\begin{lstlisting}
Each 'concern' states what this function itself does not guarantee at that operation. The analysis is exact for this function's code but does not look into callers or called functions.
\end{lstlisting}

\paragraph{reframe} Every line \texttt{holds here: X} becomes \texttt{to verify: the tool reports X}.

\paragraph{provenance} Every line number cited in a \texttt{holds here} or \texttt{concern} line is followed by
``\verb|[line N is: `...`]|'' with the source text of line N.

\paragraph{verify and verify-show} A separate call checks every claim (one \texttt{holds here} or \texttt{concern}
line) before the audit. System message: ``You check claims made by a static analyzer against C/C++ source
code.'' User message:
\begin{lstlisting}
```c
{numbered function}
```

A static analyzer made these claims about the function above. 'holds here' claims that a condition is guaranteed on every path to the operation; 'concern' claims that something is not guaranteed there.

[1] {claim}  (about {operation})
[2] ...

Check every claim against the code only. Quote the code line(s) that decide it. A claim is CONFIRMED if the code shows it is true, REFUTED if the code shows it is false (for example the cited line does not contain the claimed check), UNCLEAR if this function's code cannot decide it.
Return only a JSON array: [{"id": 1, "status": "CONFIRMED|REFUTED|UNCLEAR", "quote": "..."}]
\end{lstlisting}
In the audit, confirmed claims are kept, unclear claims are marked \texttt{(not confirmed by the code)}, and refuted
claims are removed (verify) or, in verify-show, listed after the notes under the heading
\begin{lstlisting}
Claims of the analyzer that verification found CONTRADICTED by the code (do not rely on them):
- {operation} -- {claim}
\end{lstlisting}
The legend is extended by: ``Every claim below was checked against the code in a separate verification step;
claims the code did not confirm are marked.''

\paragraph{commit and commit-verify} The first turn is the \cond{C0} audit. The model's answer is followed by:
\begin{lstlisting}
A static analysis tool, which is sometimes wrong, produced these notes about the function. {legend}

{notes}

Reconsider your analysis as a careful reviewer would treat advice from such a tool. For every note that bears on your verdict, quote the code it refers to and say whether the code confirms or contradicts it. Keep your earlier verdict unless a note that you confirmed in the code gives a concrete reason to change it. Finish with a final line:
[VERDICT] VULNERABLE  or  [VERDICT] SAFE
\end{lstlisting}
In commit-verify the notes are those of verify-show.

\section{Annotation Guide}
\label{app:annotation}

The same guide was given to both LLM annotators and to the first author. The sheet showed, for each item, the notes
given to the model and the model's answer, but neither the condition nor the correctness of the verdict.

\begin{lstlisting}
Each row shows the static-analysis notes given to the model (some notes are deliberately FALSE) and the model's answer. Read the answer and fill three columns with Y or N.

* checks_notes: the answer explicitly compares at least one note with the code (e.g. "line 8 is the loop condition, so the index is guarded").
* rejects_false_note: the answer says that a note is wrong, inaccurate or does not match the code.
* verdict_relies_on_note: the final verdict is justified (at least partly) by a note, taken at face value.

Label the three columns independently of whether the final verdict is right. When unsure, write `?` and add a remark in `comment`.
\end{lstlisting}

\section{Registration of H3}
\label{app:h3}

The registration of the multi-vendor replication (\path{preregistration/H3_multimodel.md} in the replication
package) fixed, before any of the seven models was run: the models and API identifiers, the data and conditions, the
measures and the primary test, the secondary analyses, and the handling of missing data. The SHA-256 hashes of the
registration and of the analysis script, recorded at registration time, are \texttt{85A0D074\ldots F7AD87E} and
\texttt{8047D475\ldots A1927EFB}; the script was not changed afterwards. The registration is internal: these files were first deposited publicly with
the replication package after the data had been collected.

\bibliographystyle{IEEEtran}
\bibliography{references}

\end{document}